**Inclusive Online Learning in Australia: Barriers and Enablers**

**Linda Marsden**
Young and Resilient Research Centre
Western Sydney University
ORCID: 0000-0001-5840-898X

**Luke Munn**
Digital Cultures & Societies
University of Queensland
ORCID: 0000-0002-1018-7433

**Liam Magee**
Institute for Culture and Society
Western Sydney University
ORCID: 0000-0003-2696-1064

**Matthew Ferrinda**
Catholic Education Western Australia
ORCID: 0000-0002-8592-1428

**Justin St. Pierre**
Catholic Education Western Australia

**Amanda Third**
Young and Resilient Research Centre
Western Sydney University
ORCID: 0000-0002-4437-0340

# Inclusive Online Learning in Australia: Barriers and Enablers

**Abstract**
While the pandemic highlighted the critical role technology plays in children's lives, not all Australian children have reliable access to technology. This situation exacerbates educational disadvantage for children who are already amongst our nation's most vulnerable. In this research project, we carried out a pilot project with three schools in Western Australia, conducting a series of workshops and interviews with students, parents, school staff members, and teachers. Drawing on rich empirical material, we identify key barriers and enablers for digitally inclusive online learning at the individual, interpersonal, organizational, and infrastructural levels. Of particular importance is that technology is only part of this story—an array of social, environmental, and skills "infrastructure" is needed to facilitate inclusive online learning. Building on this finding, we ran a Digital Inclusion Studio to address this holistic set of issues with strongly positive feedback from participants. We conclude with a set of recommendations for stakeholders (parents, schools, government agencies) who wish to support more digitally inclusive learning.

**Contents**

**Introduction**
The COVID-19 pandemic highlighted the critical role technology has come to play in children's education. Almost three-quarters of Australian children reported increased technology use during lockdowns (Li et al., 2022), providing them with a vital point of continuity and connection in a radically disrupted world. However, not all Australian children have routine and reliable access to technology (Thomas et al., 2020). Already among our nation's most vulnerable, these children are at high risk of compounded disadvantage, particularly as it relates to their education (Drane, Vernon and O'Shea, 2020; Preston, 2020). The pandemic underscored the educational deficits many face due to the lack of technical infrastructure and human resources needed to support home-based education.

Access to technology is fundamentally an issue of digital inclusion. Some students are empowered in their education, while others are cut off and cut out. In this article, we specifically frame digital inclusion as the ability to use digital capacities, where students can "mobilize material and symbolic resources in order to maximize benefits, opportunities and aspirations afforded by changing digital technologies and techniques" (Magee et al., 2016). This definition highlights the sociocultural, organizational, and infrastructural aspects of inclusion, while foregrounding its benefits—in this case as they relate to the educational outcomes of Australian students.

What are the factors that include some students and block others? This research aimed to identify key barriers and enablers for digitally inclusive learning-at-home. Conducted within a project of school-based technology delivery, it examines impacts of providing access to affordable Internet and appropriate devices, and of building capability to use and benefit from these technologies. By identifying barriers and enablers, the research aimed to inform scalable digital inclusion at regional and national levels.

To carry out this study, we worked with three schools in Western Australia, with high percentages of students from low-income families, with English as an Additional Language or Dialect (EALD) or indigenous backgrounds, and a self-reported lack of access to the Internet at home. We collaborated with 76 students, 22 parents, and 17 school staff across eight research workshops and seven teacher interviews. Through the workshops and interviews, we identified key barriers and enablers to digital inclusion at the individual, interpersonal, organizational and infrastructural levels. In response to the findings from the workshops and interviews, we designed and implemented a digital inclusion studio in one of the schools, which aimed to address barriers to digital inclusion in concrete ways.

We first situate our intervention in relation to recent research on remote learning and digital inclusion. The second section describes the context for the pilot study and the methods we employed. The third section describes the barriers and enablers we uncovered under individual, interpersonal, organizational and infrastructural categories. The fourth section describes our intervention in the form of the digital inclusion studio. Finally, we conclude by offering a concrete set of recommendations to a diverse set of stakeholders (schools, government, technology providers) for improving digital inclusivity.

**Literature Review**
Digital inclusion takes on increased importance in a digitally mediated world. Daily activities such as communication with friends and family, organizing finances, working, learning and managing health and wellbeing, and civic participation are increasingly dependent on the individuals ability to access and use digital technologies effectively (Thomas et al., 2020). For this reason, digital inclusion has been positioned as critical to children's rights and citizenship (Livingston and Third, 2017; Swist and Collin, 2015), to health (Sieck et al., 2021), and to social and economic participation (Helsper and Galácz, 2009). Social and digital inequalities have been demonstrated to intertwine and reinforce each other (Ragnedda et al., 2020). During the pandemic children who were already experiencing social disadvantage, such as those with disabilities, in low-income households, living in rural areas, or belonging to marginalized communities, were more likely to lack access to digital technologies, or the ability to use them effectively. These barriers risk widening the deep educational inequalities many children already face (Human Rights Watch, 2021).

The importance of digital inclusion is increasingly recognised by both public and private stakeholders. The United Nations (2020), for instance, has elevated digital inclusion to a Sustainable Development Goal, seeing it as key to harnessing digital opportunities for underserved populations. Internationally, the importance of digital inclusion to education has been recognised by governments and non-government organizations who have committed to a range of responses to address the digital divide (New Zealand Government, 2019; Vlies, 2020; Zelezny-Green et al., 2018). This recognition has only grown with the ensuing exacerbation of existing inequalities since COVID (Bocconi and Lightfoot, 2021; Bowyer et al., 2021; Dorn et al., 2020; Human Rights Watch, 2021; Ingram, 2021).

As the meaning of the 'digital' evolves, so too do methods of assessment (Magee et al., 2016). In the Australian context, a prominent multiyear study by the Australian Digital Inclusion Alliance has defined and measured digital inclusion through three dimensions: Access, Affordability, and Digital Ability (ADII, 2023). These dimensions are further operationalised as internet technology and internet data allowance (access), relative expenditure and value of expenditure (affordability), and attitudes, skills and activities (ability) (Thomas et al., 2020). While measuring access and affordability is a relatively straightforward process, defining and measuring the human aspect of digital inclusion has been widely debated, leading to the adoption of various terms such as digital competence, digital literacy, and digital capability (see Dezuanni, 2018, for a summary of this debate and list of digital ability self-assessment tools available online). While drawing upon these measures, our study explores inclusion as a set of interrelated factors, involving a mix of technological and social barriers and enablers, and we draw upon and adapt a more general socio-ecological framework in discuss these factors (Bronfenbrenner, 1977, 1989; Serrano-Santoyo and Rojas-Mendizabal, 2017).

While the Australian Digital Inclusion Index shows young people to be amongst the most digitally included age groups in Australia, qualitative research demonstrates that achieving this level of inclusion may not tell the entire story of inclusion for young people. An in-depth study of digital inclusion in young people in Australia by Third et al. (2019) identified that young people found devices expensive, dated or obsolete, and incompatible or easily broken, with costs for repairs prohibitive. Difficulties relating to connectivity include the cost and limits of data- and speed-rated pre- or post-paid plans; reliance on public wi-fi hotspots; and the unreliability of connection when on mobile or at home or school. At home, young people were sometimes reliant on devices shared with other family members, and subject to parental limits on connection times and uses. Third et al. (2019) also discovered that at school, young people must comply with device policies, limitations and restrictions on broadband access, along with inconsistent teacher attitudes and skills supporting digital use in their pedagogical practices. Other studies augment these insights, suggesting that household financial situations may also have a critical impact (Thomas et al., 2020; Preston, 2020; Graham and Sahlberg, 2021).

Integral to each model of school digital inclusion is the role of teachers and how effectively they can facilitate and model digital inclusion. However, in Australia teachers are frequently caught between competing demands of incorporating technology into their teaching practices and being conscious of e-safety issues (Wood et al., 2020). The recent emergence of generative AI has also led to renewed concerns about plagiarism, inaccuracy, legality and equity of digital content and tools in school settings (Vukovic and Russell, 2023). In Australia, ABS statistics from 2016 suggest that approximately 5% of public school students do not have fixed internet access in the home (Australian Bureau of Statistics, 2016). Preston (2020) analyzes these statistics and demonstrates that housing status (overcrowding, insecure and unsuitable), family structure (single parent), home language and English proficiency, disability, and remoteness all feed into this fixed internet access measure. The lack of fixed internet access, compounded by other factors, are barriers to young people accessing and succeeding at online education.

A raft of post-pandemic research has identified what did and did not work during lockdowns in Australia when schools delivered a learning-at-home model. Reports found that teachers and schools pivoted to a model where socio-emotional and psychological wellbeing of students was prioritized over learning outcomes (Drane et al., 2020; Flack et al., 2020; Masters et al., 2020). Educational sectors, states, and individual schools responded with differing levels of success to the challenges faced (Flack et al., 2020; Wade et al., 2023). Teachers played a critical role as the primary interface between school, home and student (Gore et al., 2020; Heffernan et al., 2021) and this came at the cost of their physical and emotional wellbeing (Learning First, 2020; Gore et al., 2020). Parents and carers were no less significant in ensuring the success of learning from home (Armitage and Loukomitis, 2020; Clinton, 2020; Parliamentary Secretary for Schools, 2020), and existing digital divides meant households from low socio-economic, EALD or indigenous backgrounds often encountered greater difficulties (Armitage and Loukomitis, 2020; Armour et al., 2020; Learning First, 2020). Resources – no or unreliable internet access – in the home environment were a factor in the success of home learning, as were the digital capabilities of students and parents particularly in primary schools (Drane et al., 2020; Learning First, 2020; Lamb et al., 2020).

These findings on student success resonate with studies in other national contexts. In Germany and Mexico, both teachers and parents played a critical role in learning from home success (Pozas et al., 2021)—and this "home environment" included both material aspects and social support from parents, friends and teachers (Dietrich et al., 2021). In Norway, critical factors included student self-efficacy, clear expectations and regular feedback from teachers, and parental support (Mælan et al., 2021). In the United States, student success was attributed to the whole family system, including caregiver wellbeing, sibling relationships, and parent-child relationships (Prime et al., 2020), while in Ireland, daily structures and routines created by parents, along with supportive family ties and peer networks, adequate equipment, private and quiet spaces to study, and individual coping mechanisms were all considered enablers of effective home learning (Émon et al., 2021).

This international context echoes some of our findings, anticipating how student success is shaped by their social environment as much as their access to digital technologies. Our research develops further evidence of barriers and enablers in the Australian post-COVID context, with particular relevance for other regional and multicultural urban settings. It also suggests how workshop-style intensives might address these systemic inequalities in modest but concrete ways to foster digital capacities.

**Context and Method**

The context for this study was three schools located in Western Australia. These schools were selected based on the number of families with a lack of internet access at home. Further factors that influenced school inclusion in the program were the number of students who were indigenous or EALD, and the income profile of families as measured by access to government benefits. One regional and two metropolitan schools were included. One of the metropolitan schools, here anonymized as "**AC** school," ran an additional educational stream for recent migrants and refugees through which students learnt English intensively. Two of the three schools operated a school-supplied device policy which was more costly for families but meant that all students had the same access to the same devices for learning. One school operated a "bring your own device" (BYOD) policy which gave families freedom to use their chosen devices but resulted in inconsistency of access to devices across the cohort.

Our pilot project with the schools aimed to directly address the three dimensions of the Australian Digital Inclusion Index: access, affordability, and ability. Devices and internet connection were co-funded by the federal government and the educational sector body as part of an investigative research project to explore barriers and enablers of digital inclusion for students who were experiencing disadvantage. Students at the school running the BYOD policy were supplied with a device ($n$ = 44) loaned to them for the school year. Twenty-five families took up the offer of 12 months free broadband access through an internet provider. In addition, a Digital Inclusion Studio (discussed further below) was held at one of the schools, aimed at enhancing the skills and capabilities of families to effectively make use of the digital technology made available to them.

The research component of the program consisted of one five-hour workshop with students in each school conducted during the school day. Separate workshops were held for English and non-English speaking students in the school delivering the intensive English education stream (4 student workshops). Additionally, a two-hour adult workshop was held in each school with a mixture of parents, carers and school staff attending. A separate workshop was held for parents with students in the intensive English educational stream (4 adult workshops). Seven one-hour teacher interviews were conducted with teachers across the three schools.

During the student workshops we worked with 76 students, 10–21 years of age, to elicit their experiences of the barriers and enablers to digital inclusion via interactive activities trialed through previous engagements with adolescent students. "Technology Map," for instance, asked children to map how technology was used in their homes, including relationships and device sharing. "Brick Wall" prompted children to identify the key barriers to digital technology they experience and what workarounds they have developed to mitigate those barriers. A separate "Postcards" activity provided an opportunity for students to share the most important issues or action-points needed to address digital exclusion.

The adult workshops (22 parents/carers, 10 staff) explored barriers and enablers to students' digital inclusion from an adult perspective, and discussed current and future impacts of exclusion. The workshop employed a socio-ecological model (discussed below), and framed activities around individual, interpersonal, organizational, community and public policy levels. The student and adult workshops were conducted in the second term of the school year on the same or consecutive days.

Teacher interviews explored the culture of technology use in the school, school expectations around use of technology in the home for learning purposes, barriers and enablers for digital inclusion, and specific digital and non-digital skills students needed for successful learning at school and home. Teacher interviews were conducted in person around the time of the workshops, or via teleconferencing in the immediate weeks following.

**Findings: Systemic Barriers and Enablers**

In reporting on the research, we employ the socio-ecological model (Bronfenbrenner, 1977, 1989; Serrano-Santoyo and Rojas-Mendizabal, 2017) to frame the factors in the student's environment when considering digital inclusion. The socio-ecological model can be conceived as a set of concentric circles, beginning from the individual child, with their biological and psychological makeup, which is impacted by their immediate physical and social environment. This microsystem, in turn, is affected by surrounding systems or mesosystems. Moving outward, the system looks to consider broader social, political, and economic conditions (exosystems) as well as the high-level beliefs and attitudes (macrosystems) maintained in a society (Bukatko and Daehler, 1998).

This model guided our identification of barriers and enablers to digital inclusion. "Barriers" are factors that appeared as obstacles to student learning, preventing them from accessing or adequately using digital technologies to carry out learning-at-home. "Enablers," by contrast, are factors that fostered digital inclusion, allowing students to embrace or adopt technologies in ways that empowered their learning. We translated the more abstract system language of the socio-ecological model to four levels: *individual*, *interpersonal*, *organizational* (encompassing schools and community) and *infrastructural* (referring to technology and governmental policy). Still viewing these levels as interconnected and mutually supporting, we cluster and present barriers and enablers according to this nested structure, beginning from the individual and expanding outward to include the interpersonal, organizational, and infrastructural. In the sections below, drawing from our engagement with students, teachers, and parents, we outline these barriers and enablers in four respective sections.

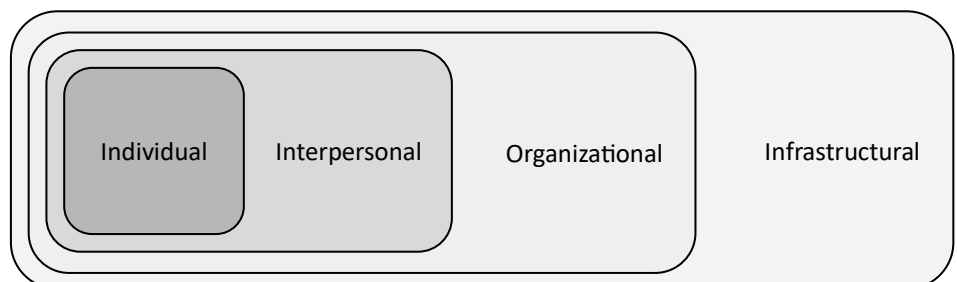

*Figure 1: Barriers and Enablers structured according to the Socio-Ecological Model*

*Individual*

Participants in our schools described a range of barriers and enablers at the individual level. For adults, learning styles and personalities (including learning difficulties or disabilities) were key factors for determining the success of learning from home. Self-control, self-discipline, self-regulation, interest, motivation, and organization were also viewed as important traits. Alongside these characteristics and abilities, a set of skills was also identified as being key for learning at home. These skills included time management, managing unsafe content and distractions online, using digital platforms effectively, critical thinking, and problem-solving.

For students, physical and psychological dispositions were mentioned often in activities. Fatigue in particular featured prominently in responses. In the "Brick Wall" activity, for example, where students identified key learning-at-home obstacles, fatigue was almost universally identified. Keywords included "fatigue," "waking up late," "sleep," "tiredness," "being tired," "sleepy," and "lack of sleep," amongst others. Other individual barriers identified included being "bored," procrastinating, lazy, (not) managing distractions, lack of drive, lack of motivation, mental blocks, learning difficulties, bad memory, and not understanding.

Students also spoke at length about the material aspects that affected their ability to learn. Smart phones and laptops were generally not shared, while tablets, desktop computers, games consoles and printers were frequently shared across household members. The **AC** cohort was an exception: students had access to a far greater number of sharing devices than the other three schools. Printers were relatively rare. As the table below shows, **A** and **AC** groups typically had larger households, with less devices per household.

| Mean # per house | **K** school | **M** school | **A** school | **AC** school |
| --- | --- | --- | --- | --- |
| people | 3.5 | 4 | 5.9 | 6.63 |
| smartphones | 3.2 | 5.2 | 4.9 | 5.7 |
| laptops | 3 | 4.5 | 2.9 | 3.2 |
| printers | 1.2 | 0.7 | 0.6 | 0.4 |

*Table 1: Mean values of people and devices per household*

The inequality of device ownership echoes student responses, which widely acknowledged these deficits were a direct result of their economic situation. Some of these aspects included devices that did not function at the required level. A lack of necessary accessories, such as mouse, keyboard, and headphones, were identified by many students, especially by those within the **AC** cohort. Headphones were repeatedly identified as a key strategic device to manage external environments which were frequently noisy and distracting.

Research identified that many home study environments presented learning obstacles. Students described working in spaces that were cramped, cluttered, noisy, poorly lit, shared/overcrowded or otherwise uncomfortable or distracting. As one student stated: "Kitchen table - not very quiet; not a lot of space; a lot of distractions; messy & dirty table; Bedroom - not motivated because I get distracted; sloppy if I work on my bed; don't have my parents around to check if I'm doing the right stuff so I get distracted" (school student, **A**).

When discussing these barriers, students often combine items that are missing in their physical, digital and pedagogical environment:

> "I sleep on the floor to do my homework. Can you please help me to get a desk and I don't have laptop, use my phone. Can I get laptop and speaker. Can you help me headphone for learn better and printer please. So I can learn better from home thanks." (student, **AC**)

> "I needed help in software like most of children in my country don't use computers at school. We have to be familiar with using different kinds of technology." (student, **AC**)
>
> "To learn better with digital technology I need better wifi so I can do my work more information about how to use technology you need mouse and desk chair and some help with translation and headphones." (student, **AC**)
>
> "I have problem with using software. Can you please take classes for us to learn. I want a better computer. Our computers are too slow." (student, **AC**)

What these responses show are connections between different elements: feelings of fatigue and sleep deprivation; lack of devices, connections and skills; and home environment – noisiness and lack of space. Students do not think about inclusion in terms of individual elements, but as interconnected links in a chain. The bullet points below summarise themes from our findings into what we refer to as *enabling chains of inclusion*:

- sleep—quiet—space—self-discipline
- bed (not floor)—desk—laptop—speaker/headphone—printer
- software—computers—different kinds of technology
- wifi—information—mouse—desk chair—translation—headphone
- software—classes—(better) computer

Many of these items are low-cost but need to be supplied together with other items in these enabling chains. Together, these statements from students, parents and teachers indicate that while learning-from-home is certainly underpinned by digital technology, successful learning hinges on several qualities—sleep, quiet times, technical know-how, adequate furniture and space for study—that are social rather than technology-dependent.

### *Interpersonal*

Students' interpersonal relationships impacted their ability to learn from home effectively, with household members, friends/peers, and teachers each playing an important role. For instance, students said that it could be difficult to discuss how to create an effective home learning environment with other household members. Young people did want their parents involved in their learning process—provided the nature of that involvement was supportive. Some students wanted their parents to support them to manage distractions. Many others appeared frustrated by what was perceived to be unhelpful limitations placed on their ability to access devices.

> [Positive Impacts]
> "I can ask for help from my parents" (student, **K**)
> "ask family to help" (student, **K**)
> "parents around so I know not do the wrong things" (student, **A**)
>
> [Negative Impacts]
> "dad interrupts my learning" (student, **M**)
> "your mum doesn't let you on it" (student, **A**)
> "parents take my phone off me; MUM!; limited screen time;" (student, **A**)

Parent workshops and teacher interviews also reiterated that carer/parents' ability to support their students' learning by creating a conducive home environment is critical. Creating this environment encompassed many factors, including parents' education level, experience with technology, attitudes towards education, ability to create boundaries, and ability to communicate with their student and school. As one teacher noted: "I think you can directly see the attitude of the parents reflected in the student's attitude." Teachers consistently identified that parents who are interested, supportive and have the skill sets themselves to model and support students enable more effective learning for students. This link between parent support and student success is echoed in other studies both in Australia (Clinton, 2020) and internationally (Émon et al., 2021; Prime et al., 2020).

Siblings also could be a source of distraction and difficulty, or conversely, of support. One student said "my little brother comes into my bedroom and annoys me while I'm doing my homework" (student, **A**), while another noted "my sister comes to help me and we do it together" (student, **A**). Friends and peers were similarly both a source of support and a barrier to learning for young people. If they were experiencing challenges at home with the task, their environment, or their equipment, they would reach out to a friend for support:

> "text my friends while doing my homework" (student, **A**)
> "if my power is cut I would go to my friend or my cousins home" (student, **AC**)
> "go to a friend's house or library" (student, **AC**)

However, students also said friends were often obstacles to learning, with "friends" appearing repeatedly in the list of obstacles to learning from home, along with "facetime," "notifications," and "social media." This points to a need for young people to be supported in learning the skills to create, maintain and respect boundaries, and to manage social distractions. Other research has also found that student's self regulation skills were important to student outcomes during lockdown (Huber and Helm, 2020; Börnert-Ringleb et al., 2021).

The attitudes and capacities of teachers to support student learning from home, along with effective communication channels that students are confident and capable of accessing, were an additional interpersonal factor for successful learning from home. For instance, students complained that they were unable to "send message to teacher" (student, **A**); that "bad teachers are quite a common problem" (student, **A**); and that there were issues with teacher communication leading to students "not understanding apps or test/assessment" (student, **A**).

Teachers also acknowledged that differing levels of enthusiasm, confidence, and aptitudes for online teaching affected experiences of learning from home. They recommended a range of measures – increased time for professional development, time to experiment and to shift learning to online modalities, and in-class support persons for the first-time new technology uses – to increase teacher confidence and capacity. For example, one teacher suggested:

> "what I would love to see is that there'll be an actual support person in their classroom. So if they are trying an activity that is based online that there would be someone there to troubleshoot (student devices around the room) if something goes wrong…a big part of that process is really giving the support in the classroom." (teacher, **A**).

This underscored the need for teachers to have support to enhance their technological and pedagogical capacities, to successfully adapt and integrate the digital into learning especially in the remote learning context (Stenman and Petterson, 2020).

According to participants, parents/carers, other household members, friends and teachers all work to enable, and occasionally inhibit, digital inclusion in education. While this iterates much longer experiences of conducive pedagogical environments generally, these comments wider recognition of the demands of digital online learning can foster improved experiences for students.

*Organizational*

At the organizational level there appeared to be several key aspects to enablement. Most importantly was equitable access to appropriate devices across a student cohort. While some of the stress on this aspect could be attributed to the study's stated focus, equitable device access was consistently identified by teachers and parents as a critical enabler of learning. The negative impacts of device inequity at the organizational level were epitomized most clearly in school **A**. This school had a "bring your own device" (BYOD) policy, forcing students who couldn't afford their own device to loan one from the school. Here, teachers described "broken devices and loan devices slow" as key barriers to learning. Given these challenges, teachers unanimously felt that device equity and uniformity were key learning enablers—especially after noticing the difference the pilot project made.

Teachers and parents both emphasized the importance of consistent use of platforms within the school. According to a **K** school parent, the communication platform used as a parent interface was "effective and allows communication between parent and school." This sentiment was echoed across other schools. However, adults noted that when platforms are not used consistently by teachers across the school, this can become a barrier to parent and student engagement. When this occurred, parents and teachers both felt frustrated and unable to adequately support students.

> "Inconsistent platforms are a barrier; consistent platforms makes parents jobs easier; effective training for teachers, students and parents in the platforms the school/sector uses" (parent/carer, **M**)
>
> "lack of clarity of expectations of school" (adult, **A**)
>
> "parents lack knowledge of technology used by students" (adult, **A**)

Teachers identified that one of their challenges through the transition to online learning was for all teachers to use school platforms consistently. In these schools this entailed using *Microsoft Teams* sites for subjects and classes, *Microsoft OneNote* for learning management, and a bespoke platform for home-school communication. Important enablers here involved sufficient teacher, student and parent training for effective platforms use. This ensured that students and parents are confident in the school's systems and processes; know how to access both materials and support as required; and contribute to the partnership between schools and families. This was a particular challenge in the schools with high percentages of EALD families, and teachers admitted sometimes struggling to communicate with parents. Many parents did not have email addresses, language or technical skills to allow them to engage in basic home school communication.

Teachers acknowledged the key role they play in learning in the learning dynamic, and stressed that their attitudes and capabilities directly impact students. They acknowledged the challenges of the rapid shift to online learning, and that they need to be prepared with suitable materials, ideally made available using the school platforms in a consistent manner. Teachers also noted that throughout the online learning phase it was challenging to ensure the content and pedagogy was appealing and suitable for their students.

> "project based learning necessarily changed to worksheets during remote learning and students found this boring; students were restricted in the materials available and this

> impacted content delivery especially for the elective classes - HPE, food tech, manual arts, arts, drama" (adult, **M**)

> "teacher preparedness - lesson materials, preparing both students and parents to effectively use; having streamlined and clear learning materials" (adult, **K**)

> "attitudes and capabilities of teachers; effective online delivery of course modules/materials; content being appealing; pedagogy suitable to online learning" (adult, **M**)

> "teachers not being skilled with technology (barrier) and more PD/training for teachers (enabler)" (adult, **A**)

Both parents and teachers, then, emphasized how important teachers were for student success. This suggests there would be great value in investing resources into enhancing teachers' ability to use digital technologies effectively.

School cultures towards digital technology are also considered crucial. Teachers acknowledged that the schools involved in this pilot study generally had very positive attitudes towards digital technology use, with broad buy-in from teaching staff. These attitudes can be helped or hindered by school and organizational policy. Adults at **A** school, for example, listed "school policies that don't support digital learning" as a barrier, and "positive school attitude for online learning" as an enabler. The Bring Your Own Device policy employed at one school, for instance, was criticized by teachers for maintaining or even exacerbating the digital divide at the school. These concrete examples demonstrate how organizational practices can cultivate digital inclusion or reinforce digital exclusion.

### *Infrastructural*

In technological terms, unreliable internet connections were one of the biggest challenges that young people faced. While most students had a home internet connection, this was consistently unreliable. Students reported being forced to turn modems and devices on and off to restore connection, wait until later, go to a friend's house or community space, hotspot via another device, or wait until they were at school in order to complete work that required online connectivity.

A key challenge for parents involved the initial activation and setup of home internet. Issues related to account sign up, installation, troubleshooting, billing, and account maintenance. They felt this was sometimes a structural or systemic issue, as internet providers could be inflexible in relation to various challenges faced by the community. Specific challenges relating to the provision of broadband included: existing lock in contracts, insufficient incentive or understanding of value of offer, and uncertainty as to what occurs at the end of the 12-month period. These linguistic, financial, contractual and structural conditions affected take-up during the program, but also illustrate challenges faced by both providers (in the supply) and community (in the acquisition) of digital devices, connectivity, and literacy.

Generally, respondents expressed concern about their uncertainty as to how and where to seek support for Internet- or digital technology-related matters, and some also noted a sense of embarrassment associated with needing such support. This could occur specifically for families from multicultural backgrounds, and addressing stigma is an important if complicated facet to which schools, families and technology providers can all contribute.

Adults acknowledged that sometimes young people utilize community spaces to access the Internet when they are having trouble connecting at home, and felt that communities should plan to incorporate free and secure wifi access in public facilities such as libraries and other public spaces conducive to learning. Students identified going to friends and cousin's houses to study as a workaround they had developed when their home environment was unsuitable.

> "I go to the library or go to friend house" (student, **AC**)

> "if my power is cut I would go to my friend/cousin's home" (student, **AC**)

> "go to the library; go to friend house; go to school" (student, **AC**)

This option emphasizes the important role of community, extended family and appropriate infrastructure to support the learning needs of young people, when studying at home is not possible.

Parents stated that there is a governmental responsibility to ensure equitable and consistent access to affordable Internet for students, which includes reliable infrastructure within urban, rural and regional areas. Participants consistently noted frustrations with certain areas considered "black spots," even in one of Australia's capital cities.

> "infrastructure reliability and access should be an equal right but continues to be inconsistent across [capital city] even for families who can afford" (adult, **M**)

> "not all Internet connections are equal" (adult, **M**)

> "ongoing challenges connecting to [the internet], ports unavailable; service providers inconsistent/unwilling to help or persist… sections of [capital city] unstable or unable to connect" (adult, **M**)

Adults also felt that it was the government's responsibility to ensure equitable access to devices for students. If education systems expect students to use devices for learning then they must ensure there are structured programs for those who are unable to afford devices. Parents also called upon the government to take the lead in ensuring that young people were safe online through regulation and governance.

**The Digital Inclusion Studio**
The insights above suggest that digital inclusion goes far beyond faster broadband and updated software. Technology is only one component in the broader environment that students need to foster a rewarding learning journey. This insight aligns with research stressing non technology-centric programs are necessary to effectively achieve sustainable digital inclusion projects, particularly in remote and underserved populations (Hatlevik and Christophersen, 2013; Marshall, 2020; Serrano-Santoyo and Rojas-Mendizabal, 2017; Zelezny-Green et al., 2018).

Based on the insights above, we modified the final workshop into what we termed a "Digital Inclusion Studio", with "studio" specifically chosen to underline the intent to create a generative, open space for instruction and experimentation. School **A** hosted the studio sessions, with a separate studio conducted for those in the intensive English program. Students from **A**, **AC** and **M** schools attended the sessions, along with carers from **AC** school and teachers from all three schools (34 students, 6 parents/carers, 5 staff members).

The studio lasted five hours, including a whole group open and closing session and seven small group sessions through which students and parents/carers rotated. Each small group session was structured to include three aspects: 1) content delivery, 2) opportunity for application or experimentation, and 3) a forum for discussion, questions, and specific expert advice.

Studios were led by an array of different stakeholders: internet providers, education sector staff, school teachers and community representatives who provided detail on migration support and services in the case of the Studio for the **AC** cohort. Topics were tailored to parent and student groups and included how to establish an ideal home-based learning space and schedule, the basics of digital devices, how to set up and maintain a home network, how to effectively use school platforms for learning and communication, and how to be safe on the internet.

Studio sessions allowed space for students and parents to obtain technical support and discuss other issues relating to their connection and study environments. This also gave the research team an opportunity to gather data, and with the support of teachers, internet providers and others, to build digital literacy through an open educational forum.

After completing rotations to each Studio, the group participated in an exit discussion during which the research team gathered data from participants on the impacts of the Studio. Students absorbed the content presented in the digital inclusion studios, were enthusiastic about their new knowledge, and readily shared their key takeaways.

Students shared their understanding of digital devices and the learning platforms their school used in several ways. First, they discussed their technical knowledge: the working parts of a laptop, how to take care of it, how to access technical support, and how to use software effectively, including text-to-speech and translate functions.

> "I have also learnt how you can do translating in OneNote." (student, **AC**)

> "I have learnt how to use shortcuts keys on devices e.g. laptops, and computers, how to use devices that have Microsoft 365 apps on it and that we need to shut down devices when not in use so that they can update throughout the day." (student, **A**)

Second, they reported that they had learned how to be safe on the internet using resources from the eSafety Commissioner's website and techniques such as blocking and reporting on their social media apps.

> "I have learnt about how you can use safety online and also how to report someone."
> "I also have to make my social media accounts private, just for my safety" (student, **AC**)

> "Stand up or stay out of the way (not get involved) of cyber-bullying" (student, **M**)

> "I learnt how to stay safe in social media and how to prevent the unsafe spam" (student, **AC**)

Third, they discussed insights about creating suitable spaces and strategies for learning at home, including study habits, scheduling and managing distractions.

> "[I've learned] to create a new schedule; to eat health food and get some rest; to not sit all day and go for a walk" (student, **AC**)

> "How you can study without getting distracted" (student, **AC**)

> "I learnt to make a schedule for my studies; I learnt that I have to take a break after 20 or 20 minutes of studying" (student, **AC**)

Likewise, parents demonstrated interest and expressed appreciation for the Studio. They reported a greater appreciation of school platforms, how to manage their home network and access resources for online safety to support their child.

> "I've learned about eSafety Commission website; how to plan for home learning; learned how *[school software system]* works and to navigate; learned about the insurance on the device loaned to my child … and how to report any damage" (parent, **A**)

> "I've learnt that our modem may be in the wrong spot; where to find parental locks [and] report abuse" (parent, **A**)

Importantly, parents and students recognised their respective roles in developing a supportive learning environment. Parents discussed how they could work collaboratively to improve learning outcomes, and where they can access further help.

> "[I've] learned the importance of discussing with your children about the use of devices/Internet while at home" (parent, **A**)

> "that we need to keep lines of communication open and flowing" (parent, **A**)

> "follow the things I learnt and tell my mom and dad what to do when you're in problems" (student, **AC**)

> "different services that help all certain types of communities" (student, **M**)

> "where and who I can go to when I need help and ways to have a more effective online work environment" (student, **A**)

Teachers were also particularly enthusiastic about the studios, commenting that there needs to be more time in their regular routines to provide this additional support to students and families to enable their digital inclusion at a level conducive to effective learning and communication.

This feedback suggests the comparatively loose and informal "carousel" format of the Digital Inclusion Studio worked well to address various student, parent/carer and teacher concerns about digital inclusion in education settings. While far from a remedy to all barriers to inclusion, the format offered opportunity for specialists to train and facilitate discussion on everything from router configuration and software use to e-safety recommendations and the role of light and sound in enabling online study. Rather than presenting students and households with a bewildering array of documentation and dedicated training options, the relaxed environment allowed participants to rotate through and "pick-and-choose" elements of a wider socio-ecology of digital inclusion at their own pace, expertise, need and level of interest.

**Conclusion**
Digital technology has become increasingly important in children's education, particularly in the pandemic and post-pandemic period. However, many children in Australia lack access to these tools and infrastructures, a condition that exacerbates existing inequalities and disadvantages. Our pilot project with three schools in Western Australia worked closely with students, parents, and teachers to identify key barriers and enablers of digital inclusion. While technical infrastructures and access certainly matter, we found barriers comprised a wide array of social and environmental factors, from noisy work environments to inconsistent parent-teacher communication. A holistic mixture of social, environmental, and skills "infrastructure" is needed to support success learning-at-home. This insight aligns with other research which underscores the importance of non-technical interventions (Hatlevik and Christophersen, 2013; Marshall, 2020). While those studies were from a development context, our research suggests digital inclusion is also a live issue in the "Global North" context of Australia. The strong positive feedback from students, parents, and teachers for the Digital Inclusion Studio activity suggests potential for fostering greater inclusion through expansion of the activity in other schools and regions.

To conclude, we offer three key areas for policy reform and further research in cultivating digital inclusion.

*Cultivating Digital Inclusion at Home*
Digital inclusion starts at home by supporting students and the parents or carers that nurture them. This support is needed by all populations, but particularly by the underserved populations like the migrant or refugee families in this research. Families need support to establish the physical aspects of a conducive home learning environment, such as access to furniture, lighting and other equipment, along with the habits and routines necessary for effective learning from home, such as scheduling, prioritizing and managing distractions. Furthermore, parents must be supported to communicate effectively with schools and teachers, including the use of any digital parent-teacher platforms. Upskilling students to harness digital technology for learning—whether this is fully using technical features or applying critical thinking and online safety to learning-at-home—is also necessary. This support should come from a variety of stakeholders, from schools to governments and educational agencies. Collective, informal and in-person events like the Digital Inclusion Studio, financed by government and delivered by local agencies, could provide this kind of support. These studios may give students and their carers the skills to navigate new technologies, to create appropriate learning environments, and to foster helpful learning habits.

*Cultivating Digital Inclusion in the Classroom*
Digital inclusion must also be fostered in the classroom, regardless of whether learning takes places remotely or in-person. Digital inclusion is a culture that needs to be actively cultivated; it emerges from a collective attitude which sees accessibility to technologies and equity of learning conditions as important, and commits to them. However, digital inclusion is also material—cultures must be underpinned by financial support, by digital devices, and by organizational commitments. This includes ensuring equitable access to functional devices for all students along with the necessary associated equipment for learning from home, such as headphones for ambient noise control. There is also a need to ensure adequate professional development time for teachers and ongoing support and space within their workload to integrate technology effectively. Furthermore, in addition to the formalized and embedded pathways for students to enhance their digital skills, such as through the Australian

Curriculum, there should be opportunities for additional support for students who need further specific support to enhance their digital inclusion. These aspects can be addressed through policy at the government level, or practically through the allocation of resources at the education sector and school level.

*Cultivating Digital Inclusion at the Internet Service Provider*
Equitable and reliable access to broadband is critical for any meaningful program of digital inclusion. This involves ensuring that access is consistent across urban, rural and regional areas, and that internet providers have flexible and responsive systems that account for the socio-cultural and socio-economic specificities of all who are seeking to access broadband. Barriers at the provider level are both technical and cultural. In our study, for instance, some migrant or refugee families found it difficult to communicate their needs to telecommunications companies, and to receive the assistance they needed to set up broadband or resolve internet issues. Governmental investment in infrastructure and regulation may need to be coupled with other policies directed towards simplification of billing systems and and staff training in intercultural competence. Community and local governments may consider investing in public spaces conducive to learning that offer free and secure wifi access along with digital inclusion support. Such interventions underscore the insight that exclusion can be embedded in powerful but subtle ways in our interfaces and structures. Digital inclusion is not just an individual's responsibility but requires systemic change.

While this study was a pilot and limited by the specificity of the education sector and the location of participating schools, we suggest that the Digital Inclusion Studio can be offered as a blueprint solution to address some of the above recommendations. The studio round-table format is flexible and provides individual and expert advice to families across the physical, financial, technological and social challenges they may be experiencing. Studio sessions can be tailored to support accessing infrastructure (setting up home internet), a home environment conducive to study, using devices and internet effectively and safely, and can provide additional wrap-around support that may be particularly beneficial to populations who are experiencing diverse challenges. In addition to immediate expert advice and building relationships at the school level, families can be provided with takeaway resources and be linked to community organizations for ongoing support. Future research should consider trialing this method in other states and educational sectors.

Our work, grounded in close collaboration with students, parents, and teachers at three schools, highlights the sociocultural alongside the technological aspects of digital inclusion. Through the statements from students, parents, and teachers above, we see a fine-grained portrait of digital exclusion emerge, one that operates in powerful but often subtle ways. This condition undermines the ability of students to learn—but as the Digital Inclusion Studio demonstrated, it can be addressed in concrete ways through crafted programs and careful interventions. Together, these insights point to formidable challenges but also highlight a series of positive improvements that could cultivate inclusive learning for the next generation of learners.

**Data Availability Statement**
The qualitative data that support the findings of this study are available, but participants have not consented to their release, and while deidentified, may contain inadvertent information that identifies participants or associates. Upon reasonable request, and within the terms of consent and permission of the authors' institutions, the data may be made available.